\documentclass[12pt,preprint]{aastex}
\usepackage{graphicx}
\begin{document}

\title{{\bf Testing theories that predict time variation of fundamental 
constants}}

\author{Susana J. Landau \altaffilmark{1}} 

\email{slandau@natura.fcaglp.unlp.edu.ar} 
\and 

\author{Hector Vucetich \altaffilmark{1}}
\email{pipi@natura.fcaglp.unlp.edu.ar}

\altaffiltext{1}{Facultad de Ciencias Astron\'{o}micas y Geof\'{\i}sicas. Universidad Nacional de La Plata. Paseo del
Bosque S/N 1900 La Plata, Argentina}

\begin{abstract}

We consider astronomical and local bounds on time variation of
fundamental constants to test some generic Kaluza-Klein-like models
and some particular cases of Beckenstein theory. Bounds on the free
parameters of the different theories are obtained. Furthermore, we
find that none of the proposed models, is able to explain recent
results \citep{Webb99,Webb00} claiming an observed variation of the
fine structure constant from quasar absorption systems at redshifts $0.5<z<3$.

\end{abstract}

\section{Introduction}

 Time variation of fundamental constants has plenty of
theoretical and experimental research since the large number
hypothesis (LNH) proposed by \citet{Dirac}. The great
predictive power of the LNH, induced a large number of research papers
and suggested new sources of variation. Among them, the attempt to
unify all fundamental interactions resulted in the development of
multidimensional theories like Kaluza-Klein
\citep{Kaluza,Klein,ChodosDetweiler,Marciano} and superstring ones
\citep{DamouryPolyakov} which predict not only energy dependence of the
fundamental constants but also dependence of their low-energy limits
on cosmological time. In such theories, the temporal variation of
fundamental constants is related with the variation of the extra
compact dimensions.

Following a different path of research, \citet{Beckenstein}
proposed a theoretical framework to study the fine structure constant
variability based on very general assumptions: covariance, gauge
invariance, causality and time-reversal invariance of electromagnetism
, as well as the idea that the Planck-Wheeler length $\left(
10^{-33}cm\right)$ is the shortest scale allowable in any theory. 

Different versions of the theories mentioned above predict different
time behaviours for the fundamental constants. Thus, experimental bounds on
the variation of fundamental constants are an important tool to check
the validity of such theories
\citep{Marciano,ChodosDetweiler,Beckenstein}.

The experimental research can be grouped into astronomical and local
methods. The latter ones include geophysical methods such as the
natural nuclear reactor that operated about $1.8\ 10^9$ in Oklo, Gabon
\citep{Oklo}, the analysis of natural long-lived $\beta$ decayers in
geological minerals and meteorites \citep{Vucetich} and laboratory
measurements such as comparisons of rates between clocks with
different atomic number \citep{Prestage}. The astronomical methods are
based mainly in the analysis of spectra form high-redshift quasar
absorption systems \citep{Drinkwater98,Webb99,Webb00,Murphy00,CyS,Bahcall}. Besides, other
constraints can be derived from primordial nucleosynthesis
\citep{nucleosintesis} and the Cosmic Microwave Background (CMB)
fluctuation spectrum \citep{Battye00,Avelino00,landau00}.

Although, most of the previous mentioned experimental data gave null
results, \citep{Webb99}, reported a significantly different
measurement of the time variation of the fine structure constant,
which was confirmed recently \citep{Webb00,Murphy00}. This suggests an
examination of the available experimental results in the context of
typical theories predicting time variation of fundamental constants.

Thus, in this work, we consider several astronomical and local bounds
on time variation of fundamental constants in the framework of two
Kaluza-Klein-like late time solutions \citep{Marciano,Bailin87} and
some particular cases of Beckenstein theory \citep{Beckenstein}. In
particular we put bounds on the free parameters of the different
models, the size of the extra dimensions in the first case, and the
parameters $l$ and $\gamma $ of Beckenstein's theory. Besides, the
consistency of experimental data with a given family of theories can be
checked.

The paper is organized as follows:\ In section II we describe briefly
the models we want to test, in section III we describe the
experimental constraints, we will use to check our models, in section
IV we present our results and briefly discuss our
conclusions.

\section{Theoretical models predicting time variation of fundamental
constants}

\subsection{Kaluza-Klein-like models}

The basic idea of Kaluza-Klein theories is to enlarge space-time to
$4+D$ dimensions in such a way that the D extra spatial dimensions
form a very small compact manifold with mean radius $R_{KK}$.

So, the metric in $4+D$ dimensions can be written :
\begin{equation}
dS^2=dt^2-r^2\left( t\right) \ g_{mn}-R_{KK}^2\left( t\right) \ g_{uv}
\label{2}
\end{equation}
where $g_{mn}$ is the metric of an $S^3$of unit radius , $r\left(
t\right) $ is the scale factor of the ordinary space, $g_{uv}$ is the
metric of an $S^D$ of unit radius and $R_{KK}\left( t\right) $ is the
scale factor of the internal space.

In Kaluza-Klein theories, gauge fields of the Standard Model of
Fundamental Interactions are related to the $g_{\mu \nu }$ elements
that connect the internal dimensions with the usual $3+1$
space-time. The gauge coupling constants are related to the
``internal'' scale of the extra dimensions through one or more scalar
fields \citep{Weinberg}.

In some models, the ``internal'' dimensions are small compared to the
large ``ordinary'' dimensions. However, at the Planck time, the
characteristic size of both internal and external dimensions are
likely to be the same. The cosmological evolution which determines the
way in which the extra dimensions are compactified depends on how many
extra dimensions are taken and on the energy-momentum tensor
considered: radiation, monopoles, cosmological constant, etc.

The generalized Einstein equations can be written as follows
\citep{KolbyTurner}:
\begin{equation}
R_{MN}=8\pi \tilde{G}\left[ T_{MN}-\frac 1{D+2}g_{MN}T_P^P-\frac
1{D+2}\frac{ \tilde{\Lambda}}{8\pi \tilde{G}}g_{MN}\right]
\label{Einstein}
\end{equation}
where $\tilde{G}$ is the gravitational constant in $4+D$ dimensions and 
$\tilde{\Lambda}$ is a cosmological constant in $4+D$ dimensions.

The evolution of the extra dimensions with cosmological time is
related with the time variation of fundamental constants through the
equation \citep{Kaluza,Klein,Marciano,Weinberg}:
\begin{equation}
\alpha _i\left( M_{KK}\right) =\frac{K_iG}{R_{KK}^2}=K_iG\,M_{KK}^2
\label{alfa}
\end{equation}
where $\alpha _i\left( M_{KK}\right) $,i=1,2,3 are the coupling
constants of $U(1)$, $SU(2)$ and $SU(3)$ for a typical energy
$R_{KK}=\frac 1{M_{KK}}$. We assume as usual, the existence of a GUT
energy scale $\Lambda_{GUT}$ beyond which all these constants merge in
only one $\alpha_i$. The $K_i$ are numbers that depend on the $D$ dimensional topology.

The expressions for the gauge coupling constants at different energies
are related through the group renormalization equation
\citep{Marciano}:
\begin{equation}
\alpha _i^{-1}\left( E_1\right) =\alpha _i^{-1}\left( E_2\right) -\frac 1\pi
\sum_jC_{ij}\left[\ln \left( \frac{E_2}{m_j}\right) +\theta \left(
E_1-m_j\right) \ln \left( \frac{m_j}{E_1}\right) \right] \label{renorm}
\end{equation}

So, we can find the low-energy limit for the gauge coupling constants
using eq.(\ref{renorm}) twice:
\begin{eqnarray}
E_1 &=&\Lambda _{GUT}\;\qquad E_2=M_{KK} \\
E_1 &=&M_W\qquad E_2=\Lambda _{GUT} \nonumber
\end{eqnarray}

Inserting eq.(\ref{alfa}) we obtain:
\begin{equation}
\alpha _1^{-1}\left( M_W\right) =\frac{K\,G}{R_{KK}^2}-\frac{76}{6\,\pi }\
\ln \left( \frac{R_{KK}^{-1}}{\Lambda _{GUT}}\right) +\frac 2\pi \ \ln
\left( \frac{\Lambda _{GUT}}{M_W}\right) \label{a1}
\end{equation}

\begin{equation}
\alpha _2^{-1}\left( M_W\right) =\frac{K\,G}{R_{KK}^2}-\frac{76}{6\,\pi }\
\ln \left( \frac{R_{KK}^{-1}}{\Lambda _{GUT}}\right) -\frac 5{3\,\pi }\ \ln
\left( \frac{\Lambda _{GUT}}{M_W}\right) \label{a2}
\end{equation}

\begin{equation}
\alpha _3^{-1}\left( M_W\right) =\frac{K\,G}{R_{KK}^2}-\frac{76}{6\,\pi }\
\ln \left( \frac{R_{KK}^{-1}}{\Lambda _{GUT}}\right) -\frac 7{2\,\pi }\ \ln
\left( \frac{\Lambda _{GUT}}{M_W}\right) \label{a3}
\end{equation}

In this way we get expressions for the gauge coupling constants
depending on $R_{KK}$ and $\Lambda _{GUT}$. In order to compare
equations (\ref{a1}), (\ref{a2}) and (\ref{a3}) with experimental and
observational values, we still should calculate the adjustment for
energies $ \sim 1$ GeV. However, since this adjustment is very small,
we will not consider it.

The gauge coupling constants are related with the fine structure
constant $ \alpha $, the QCD energy scale $\Lambda _{QCD}$ and the
Fermi coupling constant $G_F$ through the following equations:
\begin{equation}
\alpha ^{-1}\left( E\right) =\frac 52\alpha _1^{-1}\left( E\right) +\alpha
_2^{-1}\left( E\right)
\end{equation}

\begin{equation}
\Lambda _{QCD} =E\exp \left[ -\frac{2\pi }7\alpha
_3^{-1}\left( E\right) \right]
\end{equation}

\begin{equation}
G_F=\frac{\pi \ \alpha _2\left( M_W\right) }{\sqrt{2}M_W^2}
\end{equation}

It has been shown that Kaluza-Klein equation are either
non-integrable, or their solutions lack of physical interest
\citep{Amina}. However, several non-exact solutions of
eq.(\ref{Einstein}) have been analized in the literature (see
\citet{Bailin87,KolbyTurner} and references therein). 

For the purposes of this paper, though, we are interested in typical late time
solutions since the data we work with belong to times not earlier than
nucleosynthesis.
Thus, we consider models where the scale factor of the Universe
behaves as in a flat Robertson-Walker space-time with and without
cosmological constant and the radius of the internal dimensions
behaves as the following schematic solutions motivated in
 \citet{Marciano,Bailin87}:
\begin{equation}
R_{KK}\left( t\right) \sim R_0+\Delta R\,\ \left(1 - \cos \left[ \omega \left(
t-t_0\right) \right]\right) \label{rkk}
\end{equation}
\begin{equation}
R_{KK}\left( t\right) \sim R_0+\Delta R\,\ \ {\left[\frac{t_0}{t}\right]}^{3/4} \label{rkk2}
\end{equation}
where $R_0=R_{KK}\left( t_{Planck}\right) \simeq R_{Planck}$. We
expect that typical solutions of Kaluza-Klein cosmologies behave
asymptotically like eqs.(\ref{rkk}) and (\ref{rkk2}) with $\Delta
R<<R_0$ and $\omega$ depending on the details of the model. We will
refer to solution \ref{rkk} as generic model 1 and to solution
\ref{rkk2} as generic model 2.
Generic model 1 is similar in shape to the variation in $\alpha$ reported by \citet{Webb00}. Indeed, it predicts a null variation of the fine structure constant today and a greater variation in the past.



Thus, the free parameter in all Kaluza-Klein-like models will be :
$\frac{\Delta R}{R_0}\ $ and we will take as usual $\Lambda_{GUT}=10^{16} GeV$.
Table \ref{tabla3} shows the
cosmological model and the values of $\omega $ 
considered for each particular model.

\subsection{Beckenstein models}

As we have mentioned above, \citet{Beckenstein} proposed a
framework for the fine structure constant $\alpha $ variability based
on very general assumptions such us: covariance, gauge invariance,
causality and time-reversal invariance of electromagnetism , as well
as the idea that the Planck-Wheeler length $\left( 10^{-33}cm\right) $
is the shortest scale allowable in any theory.

He obtained the following equation for the temporal variation of
$\alpha $:
\begin{equation}
\left( \frac{a^3\dot{\varepsilon}}\varepsilon \right) ^{.}=-a\left( t\right)
^3\varsigma \left( \frac{l^2}{\hbar \ c}\right) \rho _mc^4
\end{equation}
where $\varepsilon =\left( \frac \alpha {\alpha _{today}}\right)
^{\frac 12}$, $l$ is a length scale of the theory, $\rho _m$ is the
total rest mass density of matter, $a\left( t\right) $ is the
expansion scale factor and $ \varsigma $ is a dimensionless parameter
which measures the fraction of mass in the form of Coulomb energy of
an average nucleon, compared to the free proton mass
(\citet{Beckenstein} assumed that $\varsigma $ is constant
and equal to $1.3\times 10^{-2}$).

In an expanding Universe where $\rho _m=\frac{3H_0^2}{8\pi G}\left[
\frac{ a\left( t_0\right) }{a\left( t\right) }\right] ^3$, we obtain:
\begin{equation}
\frac{\dot{\varepsilon}}{\varepsilon} =-\varsigma \left(
\frac{l^2 c^3}\hbar \right) \rho _m\left( t-t_c\right) \label{beck}
\end{equation}
where $t_c$ is an integration constant. We consider a flat model with
cosmological constant where the scale factor varies as:
\begin{equation}
a\left( t\right) =a\left( t_0\right) \left( \frac{\Omega _m}{\Omega _\Lambda 
}\right) ^{\frac 13}\left[ \sinh \left( \frac 32\Omega _\Lambda
^{1/2}H_0t\right) \right] ^{\frac 23}
\end{equation}

 Integrating eq.(\ref{beck}), we obtain the time variation of the
fine structure constant as follows:
\begin{equation}
\frac{\Delta \alpha }\alpha =-\frac{3\ \varsigma }{8\ \pi }\left(
H_0t_0^{-1}\right) ^2\;\left( \frac l{L_p}\right) ^2\left[ 
\begin{array}{c}
\beta \coth \beta -%
{\displaystyle {t \over t_0}}
\beta \coth \left( \beta \frac{t}{t_0}\right) +\ln \left( \frac{\sinh \left(
\beta \frac{t}{t_0}\right) }{\sinh \left( \beta \right) }\right) \\ 
+\gamma \left( \beta \coth \left( \beta \frac{t}{t_0}\right) -\beta \coth
\beta \right)
\end{array}
\right]
\end{equation}
with
\[
\coth \beta =\Omega _\Lambda ^{-\frac 12} 
\]
where $L_p=\left( \frac{G\hbar }{c^3}\right)
^{\frac 12}.$
In all cases the integration constant is such that $\varepsilon \left(
t_0\right) =1$ and $\Omega _m+\Omega _\Lambda =1$

Table \ref{tabla4} shows the cosmological parameters for the models we use
to test this theory. 
The free parameters in this models are $L=\frac l{L_p}$ and $\gamma
$. 

\section{Bounds from astronomical and geophysical data}

In this section, we make a critical discussion of the rather heterogeneous
data set we use to test our models.

\subsection{The Oklo Phenomenon}

One of the most stringent limits on time variation of fundamental
constants follows from an analysis of isotope ratios of
$^{149}\rm{Sm}/^{147}\rm{Sm}$ in the natural uranium fission reactor
that operated  $1.8\times 10^9$ yr ago at the present day site of the Oklo mine
in Gabon, Africa \citep{oklo2,Oklo}. From an analysis of nuclear and
geochemical data, the operating conditions of the reactor could be
reconstructed and the thermal neutron capture cross sections of
several nuclear species measured. In particular, a shift in the lowest
lying resonance level in $^{149}{\rm Sm}: \Delta = E_r^{149{\rm(Oklo)}} -
E_r^{149{\rm(now)}}$ can be derived from a shift in the neutron capture
cross section of the same nucleus \citep{oklo2,Oklo}. 
We know that we can translate the shift in $\Delta $ into a bound on a
possible difference between the values of $\alpha $ and $G_F$ during
the Oklo phenomenon and their value now.  \citet{Oklo} derived 
bounds on $ \alpha $ and $G_F$   separately; here we
consider both variations at the same time as follows:
\begin{equation}
\Delta =\alpha \frac{\partial E_r}{\partial \alpha }\frac{\Delta \alpha }
\alpha +G_F\frac{\partial E_r}{\partial G_F}\frac{\Delta G_F}{G_F}
\end{equation}
where $\Delta \alpha = \alpha ^{Oklo}-\alpha ^{now}$ and $\Delta
G_F=G_F^{Oklo}-G_F^{now}$. The value of $\Delta $ is shown in Table
\ref{tabla1}. Finally, using the values of $\Delta, \alpha \frac{
\partial E_r}{\partial \alpha },G_F\frac{\partial E_r}{\partial G_F}$
from \citet{Oklo}, we can relate $\Delta$ with $\frac{\Delta
\alpha}{\alpha}$ and $\frac{\Delta G_F}{G_F}$ (see first entry in
Table \ref{tabla2}).

\subsection{Long-lived $\beta $ decayers}

The half-life of long-lived $\beta $ decayers such $^{187}\rm{Re}, ^{40}\rm{K}, ^{87}\rm{Rb}$
has been determined either in laboratory measurements or by comparison with
the age of meteorites, as found from $\alpha $ decay radioactivity analysis.
\citet{Vucetich} have derived a relation between the
shift in the half-life of three long lived $\beta $ decayers and a possible
variation between the values of the fundamental constants $\alpha ,\Lambda
_{QCD}$ and $G_F$ at the age of the meteorites and their value now (see
entries 2,3 and 4 of Table \ref{tabla2}).

The values of $\frac{\Delta \lambda }\lambda $ for $^{187}\rm{Re}$ ,
$^{40}\rm{K}$, $^{87}\rm{Rb}$ are respectively shown in entries 2, 3, and 4 in
Table \ref {tabla1} where $\Delta =\frac{\Delta \lambda }\lambda $ and
$\Delta \lambda =\lambda (t=5.535\times 10^9)-\lambda \left(
t=t_0=1.0035\times 10^{10}\right) .$

\subsection{Laboratory experiments}

The best limit on $\alpha $ variation, comes from a laboratory
experiment \citep{Prestage}; it is a limit on a present day variation
of $\alpha $.
The experiment is based on a comparison of rates between clocks based
on hyperfine transitions in atoms with different atomic number
. H-maser and Hg+ clocks have a different dependence on $\alpha $
since their relativistic contributions are of order $\left( \alpha
Z\right) ^2$. The result of a 140 day clock day comparison between an
ultrastable frequency standard based on Hg+ ions confined to a linear
ion trap and a cavity tuned H maser \citep{Prestage} is shown in Table \ref{tabla1} where $\Delta =\frac{ \Delta \alpha
}\alpha $.

\subsection{Quasar absorption systems}

Quasar absorption systems present ideal laboratories  to
search for any temporal variation in the fundamental constants. The
continuum spectrum of a quasar was formed at an epoch corresponding to
the redshift $z$ of main emission details specified by the
relationship $ \lambda _{obs}=\lambda _{lab}\left( 1+z\right) $.
Quasar spectra of high redshift show the absorption resonance lines of
the alkaline ions like CIV, MgII, FeII, SiIV and others, corresponding
to the $ S_{1/2}\rightarrow P_{3/2}\left( \lambda _1\right) $ and
$S_{1/2}\rightarrow P_{1/2}\left( \lambda _2\right) $ transitions. The
relative magnitude of the fine splitting of the corresponding
resonance lines is proportional to the square of the fine structure
constant $\alpha $ to lowest order in $\alpha $.
\begin{equation}
\frac{\Delta \lambda }\lambda =\frac{\lambda _1-\lambda _2}\lambda \sim
\alpha ^2
\end{equation}

Therefore, any change in $\alpha $ will result in a corresponding
change in $ \Delta \lambda $ in the separation of the doublets of the
quasar as follows:
\[
\frac{\Delta \alpha }\alpha =%
{\textstyle {1 \over 2}}
\left[ \frac{\left( \frac{\Delta \lambda }\lambda \right) _z}{\left( \frac{%
\Delta \lambda }\lambda \right) _{now}}-1\right] 
\]

\citet{CyS}, \citet{Varshalovich} and \citet{Murphy01} have applied
this method to SiIV 
doublet absorption lines systems at different redshifts ($2.5 < z < 3.33$)
to find the values shown in entries 6 to 10 of table \ref{tabla1}
where $\Delta =\frac{ \Delta \alpha }\alpha .$

\citet{Webb99} have improved this method comparing
transitions of different species, with widely differing atomic masses. As
mentioned before, this is the only data consistent with a time varying
fine structure constant. In turn, recent work \citep{Webb00,Murphy00}
including new optical data  
confirms their previous results. The values of
$\frac{\Delta \alpha } \alpha $ at redshift $z=1.2$, $z=2.7$ and $z=2.5$ are
respectively shown in entries 11, 12 and 13 of Table \ref{tabla1}.

Moreover, the ratio of frequencies of the hyperfine 21 cm absorption
transition of neutral hydrogen $\nu _a$ to an optical resonance
transition $ \nu _b$ is proportional to $x=\alpha ^2g_p\frac{me}{mp}$
where $g_p$ is the proton $g$ factor. Thus, a
change of this quantity will result in a difference in the redshift
measured from 21 cm and optical absorption lines as follows:
\begin{equation}
\frac{\Delta x}x=\frac{z_{opt}-z_{21}}{\left( 1+z\right) }
\end{equation}
So, combining the measurements of optical and radio redshift, a bound on $x$
can be obtained.

The upper bounds on $x$ obtained by \citet{CyS} at redshift $z=1.776$
are shown in Table \ref{tabla1} where $\Delta = \frac{\Delta x}x$. The
relationship between $\frac{\Delta x}x$ and the variation of $\alpha
$, $G_F$ and $ \Lambda _{QCD}$ is shown in table \ref{tabla2}.  Other
bounds on $x$ were obtained by \citet{WolfeyDavis} at redshift $
z=0.69$ (entry 15 of Table \ref{tabla1}) and \citet{WolfeyBrown} at
redshift $z=0.52$ (entry 16 of Table \ref{tabla1})

On the other hand, the ratio of the rotational transition frequencies of
diatomic molecules such as CO to the 21 cm hyperfine transition in
hydrogen is proportional to $y=g_p\alpha ^2$. Thus, any variation in
$y$ would  be observed as a difference in the 
redshifts measured from 21 cm and molecular transition lines:
\begin{equation}
\frac{\Delta y}y=\frac{z_{mol}-z_{21}}{\left( 1+z\right) }
\end{equation}

\citet{Murphy02} have placed upper limit on $y$ at
redshift $ z=0.25$ and at redshift $z=0.68$. The observed values are
shown in entries 17 and 18 of Table \ref{tabla1}, where $\Delta
=\frac{\Delta y}y$. Entries 17 and 18 of Table \ref{tabla2} relate
$\frac{\Delta y}y$ with the variation of $\alpha $.

Finally, observations of molecular hydrogen in
quasar absorption systems can be used to set bounds on the evolution
of $\mu=\frac{m_e}{m_p}$.
The most stringent bounds established by \citet{pothekin98} are shown in entry 19 of Table \ref{tabla2}.

\subsection{Nucleosynthesis}

Primordial nucleosynthesis also provides a bound on the variation of
fundamental constants. A didactical analysis of $^4$He production can be
found in \citet{nucleosintesis}. At the conclusion of the big-bang
nucleosynthesis the $^4$He mass fraction of the total baryonic mass is
given by \citep{nucleosintesis}: 
\begin{equation}
Y=2\exp [-\frac{t_c}\tau ]\ X\left( t_F\right) \label{helio}
\end{equation}
where $t_c$ is the neutron capture time, $\tau $ is the neutron mean life
and $X\left( t_F\right) $ is ratio of the neutron to total baryon number at
the time where the baryons become uncoupled from the leptons (freeze-out 
time).

In appendix I, we derive the following expression for the change in the
helium abundance $\Delta Y$ brought about by changes in the fundamental
constants:
\begin{equation}
\frac{\Delta Y}Y=0.74\frac{\Delta R_{KK}}{R_{KK}}+0.64\frac{\Delta
G_F}{G_F}+1.76\frac{\Delta \alpha }\alpha -0.3\frac{\Delta \Lambda
_{QCD}}{\Lambda _{QCD}}
\end{equation}

\subsection{Cosmic Microwave Background}

Any variation of the fine structure constant $\alpha$ alters the
physical conditions at recombination and therefore changes the cosmic
microwave background (CMB) fluctuation spectrum. Moreover, the
fluctuacion spectrum of CMB is sensitive to many cosmological
parameters such as the density of barionic and dark matter, the Hubble
constant and the index of primordial spectral fluctuations. Recently,
different independent analysis \citep{Battye00,Avelino00,landau00}
showed that the recent published data of Boomerang and Maxima are
better fitted with a varying fine structure constant and a density of
baryonic matter closer to nucleosynthesis bounds. The same authors established a bound on 
$\alpha $ variation at the epoch at which neutral hydrogen formed (see entry 21 in Table \ref{tabla1}).

\section{Results and Discussion}

From the data rewiewed in the last section, we have performed a statistical
analysis working on $\chi ^2$ function with MINUIT to compute the
best-fit parameter values and uncertainties including correlations
between parameters.


For the Kaluza-Klein like models, results within $99\%$ of confidence
level $\left( 3\sigma \right) $ are shown in table \ref{tabla3}. For
the models derived from Beckenstein's proposal we obtain results with
$ 90\%$ of confidence level (see table \ref{tabla4}).  The contours of
the likelihood functions for Beckenstein's models in regions of 90 \%
and 70 \% of confidence level are shown in figures 1 and 2.

The values of the free parameters obtained are coincident within
uncertainties for the Kaluza-Klein like models (table\ref{tabla3}) and
for Beckenstein's models (table \ref{tabla4}).  Besides, the values
obtained are consistent with theoretical supposition $\Delta R<<R_0$
for Kaluza-Klein like models, but they disagree with the supposition
$l>L_p$ implied in Beckenstein's framework.

Thus, the present available data set, considered within Bekenstein's
framework, is capable to rule out $\alpha$ variability, while the
original paper had to recourse to E\"otv\"os-like experiments to
achieve the same result.  \citet{LyS} have also analyzed
$\alpha$ variation in the context of Bekenstein's theory. Our results
are in agreement with their analysis, even though they didn't allow
both free parameters of the theory: $\frac{l}{L_p}$ and $\gamma$ to
vary independently.

However, it should be noted that Beckenstein's framework is very
similar to the dilatonic sector of string theory, and it has been
pointed out that in the context of string theories
\citep{strings1,strings2} there is no need for an universal relation
between the Planck and the string scale.

Finally, our results are consistent with no time variation of
fundamental constants over cosmological time in agreement most of the
experimental results. Indeed, excluding the Webb et al. data points
from our fits does not change significantly the values of the adjusted
constants. Thus, this rather large class of theories cannot explain
this discrepant result.

\acknowledgements
The authors whishes to thank Professor D. Harari for many interesting
discussions. H. V. acknowledges economic support from grant G035-UNLP.

\appendix
\section{Appendix I}

Following \citep{nucleosintesis} and eq. \ref{helio}, the change in the
helium abundance is given by:
\begin{equation}
\frac{\Delta Y}Y=\frac{t_c}\tau \left( \frac{\Delta \tau }\tau -\frac{\Delta
t_c}{t_c}\right) +\frac{\Delta X\left( t_F\right) }{X\left( t_F\right) }
\label{1}
\end{equation}
where
\begin{equation}
\frac{\Delta X\left( t_F\right) }{X\left( t_F\right) }=-0.52\frac{\Delta b}b
\end{equation}
and 
\begin{equation}
b=255\left( \frac{45}{4\pi N}\right) ^{1/2}\frac{M_{pl}}{\tau \ Q^2}
\end{equation}

\begin{equation}
Q=\Delta m=m_n-m_p
\end{equation}
where $N$ is the number of neutrino types.

Since $\tau =Q^5G_F^2$, we find for the ratio of neutron to total baryon
number at the freeze-out time:
\begin{equation}
\frac{\Delta X\left( t_F\right) }{X\left( t_F\right) }=-0.52\left[
\frac{ \Delta M_{pl}}{M_{pl}}-2\frac{\Delta G_F}{G_F}-7\frac{\Delta
Q}Q\right]
\label{2bis}
\end{equation}

Next, also from \citep{nucleosintesis} we take the following expression
for the neutron time capture:
\begin{equation}
t_c=\left( \frac{45}{16\pi N}\right) ^{1/2}\left( \frac{11}4\right)
^{2/3} \frac{M_{pl}}{T_{\gamma ,c}^2}+t_0
\end{equation}
where $t_0$ is an integration constant, $T_{\gamma ,c}$ is the temperature
of the photon at the neutron capture time. Thus, the last equation yields:
\begin{equation}
\frac{\Delta t_c}{t_c}=\frac{\Delta M_{pl}}{M_{pl}}-2\frac{\Delta T_{\gamma
,c}}{T_{\gamma ,c}} \label{3}
\end{equation}

Writing $T_{\gamma ,c}=\frac{\varepsilon _D}{z_c}$ with $\varepsilon _D=m_n+m_p-m_D$ and $z_c=\frac{\varepsilon _D}{T_{\gamma ,c}}$ we obtain:

\begin{equation}
\frac{\Delta T_{\gamma ,c}}{T_{\gamma ,c}}=\frac{\Delta \varepsilon
_D}{ \varepsilon _D}-\frac{\Delta z_c}{z_c}=\frac{\Delta \Lambda_{QCD}}{\Lambda _{QCD}}-\frac{\Delta z_c}{z_c} \label{4}
\end{equation}

Since at the neutron capture time, the neutrons are essentially all
converted into helium, we may identify the temperature $T_{\gamma ,c}$
at which neutrons are captured, or equivalently the redshift
$z_c=\frac{ \varepsilon _D}{T_{\gamma ,c}}$ , by the condition:
\begin{equation}
\left( \frac{dX_D}{dz}\right) _{z=z_c}=0
\end{equation}
where $X_D$ is the ratio of deuterons to total baryon number.

From \citep{nucleosintesis} it is easy to see that the last equation is
 equivalent to the following:
\begin{equation}
f\left( z_c\right) =\ln \left( C_0\right) +%
{\textstyle {4 \over 3}}
\ln \left( \frac{\varepsilon _D}{m_p}\right) +\ln \left( \frac{M_{pl}}{m_p}%
\right) ^{}+%
{\textstyle {4 \over 3}}
\ln \left( \alpha \right) -%
{\textstyle {17 \over 6}}
\ln \left( z_c\right) +z_c-5.11\frac{\alpha ^{\frac 12}z^{\frac 13}}{\left( 
\frac{\varepsilon _D}{m_p}\right) ^{\frac 13}}=0
\end{equation}
where $C_0$ is a constant and $z_c=26$.

Assuming:
\begin{equation}
\delta f=\left( \frac{\partial f}{\partial z}\right) _{z=z_c}^{\alpha
_i=\alpha _{io}}\delta z+\left( \frac{\partial f}{\partial \alpha
}\right) _{z=z_c}^{\alpha _i=\alpha _{io}}\delta \alpha +\left(
\frac{\partial f}{ \partial M_{pl}}\right) _{z=z_c}^{\alpha _i=\alpha
_{io}}\delta M_{pl}+\left( \frac{\partial f}{\partial \varepsilon
_D}\right) _{z=z_c}^{\alpha _i=\alpha _{io}}\delta \varepsilon _D=0
\end{equation}
where $\alpha _i=\alpha _{io}$ means $\alpha =\alpha _{today}$ and
$\Lambda _{QCD}=\Lambda _{QCDtoday}$ we obtain the following
expression:
\begin{equation}
\frac{\Delta z_c}{z_c}=-\left[ \left( \frac{\partial f}{\partial
\alpha } \frac \alpha z\right) _{z=z_c}^{\alpha _i=\alpha
_{io}}\frac{\Delta \alpha } \alpha +\left( \frac{\partial f}{\partial
M_{pl}}\frac{M_{pl}}z\right) _{z=z_c}^{\alpha _i=\alpha
_{io}}\frac{\Delta M_{pl}}{M_{pl}}+\left( \frac{ \partial f}{\partial
\varepsilon _D}\frac{\varepsilon _D}z\right) _{z=z_c}^{\alpha
_i=\alpha _{io}}\frac{\Delta \varepsilon _D}{\varepsilon _D} \right]
\left( \frac{\partial f}{\partial z}\right) ^{-1} \label{zeta}
\end{equation}

Evaluating eq.(\ref{zeta}) yields:
\begin{equation}
\frac{\Delta z_c}{z_c}=-0.13\frac{\Delta \alpha }\alpha +0.046\frac{\Delta
M_{pl}}{M_{pl}}+0.26\frac{\Delta \Lambda _{QCD}}{\Lambda _{QCD}} \label{5}
\end{equation}

Thus, from eqs. (\ref{1}), (\ref{2bis}), (\ref{3}), (\ref{4}),
(\ref{5}) and as $\frac{ \Delta Q}Q=\frac{\Delta \alpha }\alpha $, the
final expression yields:
\begin{equation}
\frac{\Delta Y}Y=0.74\frac{\Delta R_{KK}}{R_{KK}}+0.64\frac{\Delta
G_F}{G_F} +1.76\;\frac{\Delta \alpha }\alpha -0.3\frac{\Delta \Lambda
_{QCD}}{\Lambda _{QCD}}
\end{equation}
where we have used the equality $R_{KK}\left( t_{pl}\right) \simeq
R_{pl}= \frac 1{M_{pl}}$



\clearpage

\begin{table}[tbp]
\caption{Observational Data. The columns show the data number
(correlated with the respective equation in Table \ref{tabla2}), the
method considered, the time 
interval for which the variation was measured in units of $10^9$ yr,
computed for models with and without cosmological constant, the observed
value, the standart deviation and the corresponding reference}
\label{tabla1}

\begin{center}
\begin{tabular}{ccccccc}
\hline
\hline
& Method & $t-t_0$ & $t-t_0$ & $\Delta $ & $\sigma \left( \Delta \right) $ & Ref. \\
& & $ \Omega_{\Lambda}=0$ & $\Omega_{\Lambda}=0.75$ & $\times 10^{-6}$ & $\times 10^{-6}$ & \\
\hline

1 & Oklo reactor & $1.8$ & $1.8$ & $-15000$ & $1050000$ & 1\\ 
2 & Long lived $\beta$ decayers & $4.5$ & $4.5$ & $0$ & $6700$ & 2 \\ 
3 & Long lived $\beta$ decayers& $4.5$ & $4.5$ & $0$ & $13000$ & 2 \\ 
4 & Long lived $\beta$ decayers & $4.5$ & $4.5$ & $0$ & $13000$ & 2\\ 
5 & Laboratory bounds & $3.8\times 10^{-10}$ & $3.8\times 10^{-10}$ & $0$ & 
$10^{-8}$ & 3 \\ 
6 & Quasar absorption systems & $8.7$ & $13$ & $0$ & $350$ & 4 \\ 
7 & Quasar absorption systems& $8.9$ & $13$ & $0$ & $350$ & 4 \\
8 & Quasar absorption systems& $8.7$ & $12.8$& $0$ & $83$ &5 \\
9 & Quasar absorption systems& $8.68$ & $12.5$ & $0$ & $80$ & 5 \\
10& Quasar absorption systems& $8.51$ & $12.24$ & $-5$ & $13$ & 6 \\
11 & Quasar absorption systems& $6.8$ & $9.17$ & $-7$ & $2.3$ & 7 \\ 
12 & Quasar absorption systems& $8.6$ & $$12.4 & $-7.6$ & $2.8 $& 7 \\
13 & Quasar absorption systems& $6.5$ & $$8.5& $-5$ & $1.3 $& 7 \\
15 & Quasar absorption systems& $7.8$ & $11$ & $7$ & $11$ & 4\\ 
15 & Quasar absorption systems& $5.5$ & $6.9$ & $0$ & $120$ & 8\\ 
16 & Quasar absorption systems& $4.7$ & $5.7$ & $0$ & $280$ & 9\\ 
17 & Quasar absorption systems &$2.9$ & $3.2$ & $-2$ & $4.4$ & 10\\ 
18 & Quasar absorption systems & $5.4$ & $6.8$ & $-1.6 $ & $5.4$ & 10\\ 
19 & Quasar absorption systems & $ 8.65$ & $12.6$ & $0 $ & $20$ & 11 \\
20 & Nucleosynthesis & $10$ & $15 $ & $0$ & $43000$ & 12\\
21 & CMB & $10 $ & $15 $ & $ 0 $ & $ 10000$ & 13,14,15 \\
\hline
\hline
\end{tabular}
\end{center}
\tablerefs{
(1) Damour and Dyson 1996;(2) Sisterna and Vucetich 1990; (3) Prestage, Toelker and Maleki 1995; (4) Cowie and Songaila
1995; (5) Varshalovich, Panchuk and Invanchik 1996; (6) Murphy et al 2001b ;Webb et al. 2000; (8) Wolfe and Davis 1979, (9) Wolfe, Brown and Roberts 1976; (10) Murphy et al 2001c ;(11) Pothekhin et al. 1998 ;(12)
Bernstein, Brown and Feinberg 1988;(13)Battye, Crittenden and Weller 2001;(15) Avelino et al. 2000;(15) Landau, Harari
and Zaldarriaga 2001}
\end{table}

\clearpage

\begin{table}[tbp]
\caption{The equation: $\Delta = a\frac{\Delta \alpha }\alpha +b\frac{\Delta G_F}{G_F}+c\frac{\Delta \Lambda_{QCD}}{\Lambda _{QCD}}$ relates the observed value ($\Delta$ of table \ref{tabla1}) with the relative variation of fundamental constants. In this table we show the coefficients of this equation for each data considered in table \ref{tabla1}}
\label{tabla2}
\begin{center}
\begin{tabular}{lllllll}
\hline
\hline
& & $ a $ & & $ b $ & & $ c $ \\ \hline
1 & & $10^6$ & & $5.6$ & & $0$ \\ 
2 & & $2.16 \times 10^4$ & & $2$ & & $5.62 \times 10^3$ \\ 
3 & & $4.6 \times 10$ & & $2$ & & $1.7 \times 10$ \\ 
4 & & $1.07 \times 10^3$ & & $2$ & & $2.71$ \\ 
5 & & $1$ & & $0$ & & $0$ \\ 
6 & & $1$ & & $0$ & & $0$ \\ 
7 & & $1$ & & $0$ & & $0$ \\ 
8 & & $1$ & & $0$ & & $0$ \\
9 & & $1$ & & $0$ & & $0$ \\
10& & $1$ & & $0$ & & $0$ \\
11 & & $1$ & & $0$ & & $0$ \\ 
12 & & $1$ & & $0$ & & $0$ \\
13 & & $1$ & & $0$ & & $0$ \\ 
14 & & $2$ & & $0$ & & $-1$ \\ 
15 & & $2$ & & $0$ & & $-1$ \\ 
16 & & $2$ & & $0$ & & $-1$ \\ 
17 & & $2$ & & $0$ & & $0$ \\ 
18 & & $2$ & & $0$ & &$0$ \\
19 & & $0$ & & $0$ & &$-1$ \\
20 & & $1.76$ & & $ 0.64 $ & & $ -0.3$ \\
21 & & $1$ & & $0$ & & $0$ \\
\hline
\hline
\end{tabular}
\end{center}
\end{table}

\clearpage

\begin{table}[tbp]
\caption{Results for the Kaluza-Klein like models. The columns show the
number of particular model considered
, the number of generic model, the cosmological
parameters and the free parameters of the theory taken as constant in
this work, the best fit parameter value and standart deviation in
units of $10^{-14}$ . $t_{01}=1.0\times 10^{10}\ $ yr is the age of the
universe for models without cosmological constant; $t_{02}=1.5\times
10^{10}\ $yr is the age of the universe for models with cosmological
constant. For all models$\ H_0=65$ $km\times seg^{-1}\times
Mpc^{-1}$}
\label{tabla3}
\begin{center}
\par
\begin{tabular}{lllllll}
\hline
\hline
& & $\Omega_m $ & $\Omega_\Lambda$ & $\omega $ & $\frac{\Delta R_{KK}}{R_{KK}}$ \\
\hline
1 & 1 & $1 $ & $0 $ & $\frac{2\pi }{t_{01}}$ & $(1.0 \pm 6.0) \times 10^{-8} $ \\ 2 & 1
& $0.25 $ & $0.75 $ & $\frac{2\pi }{t_{02}}$ & $(2.1 \pm 8.4) \times 10^{-8}$ \\ 3 & 2
& $1 $ & $0 $ & $ - -$ & $3 \times 10^{-19} \pm 2 \times 10^{-16} $
\\ 4 & 2 & $0.25$ & $0.75 $ & $ - - $ & $2.5 \times 10^{-18} \pm 9
\times 10^{-15}$ \\
\hline
\hline
\end{tabular}
\end{center}
\end{table}

\begin{table}[tbp]
\caption{Results for the Beckenstein's type models. The columns show
the number of particular model, the cosmological 
parameters, the value and standard deviation of the best fit
parameters and the correlation coefficient. For 
all models $H_0=65$ $km\times seg^{-1}\times Mpc^{-1}$}

\label{tabla4}
\begin{center}
\par
\begin{tabular}{llllllll}
\hline
\hline
& $\Omega _m $ & $\Omega _\Lambda$ & $L$ & $\gamma $ &
$\rho\left(L,\gamma\right)$\\
\hline
$1$ & $1$ & $0$ & $0.0021^{+0.018}_{-0.0011}$ & $252^{+110}_{-90}$ &
$-0.001$ \\ 
$2$ & $0.25$ & $0.75$ & $ {10^{-5}}^{+0.0003}_{-0.8 \times 10^{-5}}$ &
$77^{+36}_{-24}$ & $-0.024$ \\
\hline
\hline
\end{tabular}
\end{center}
\end{table}

\clearpage

\begin{figure}
\begin{center}
\includegraphics[scale=0.4]{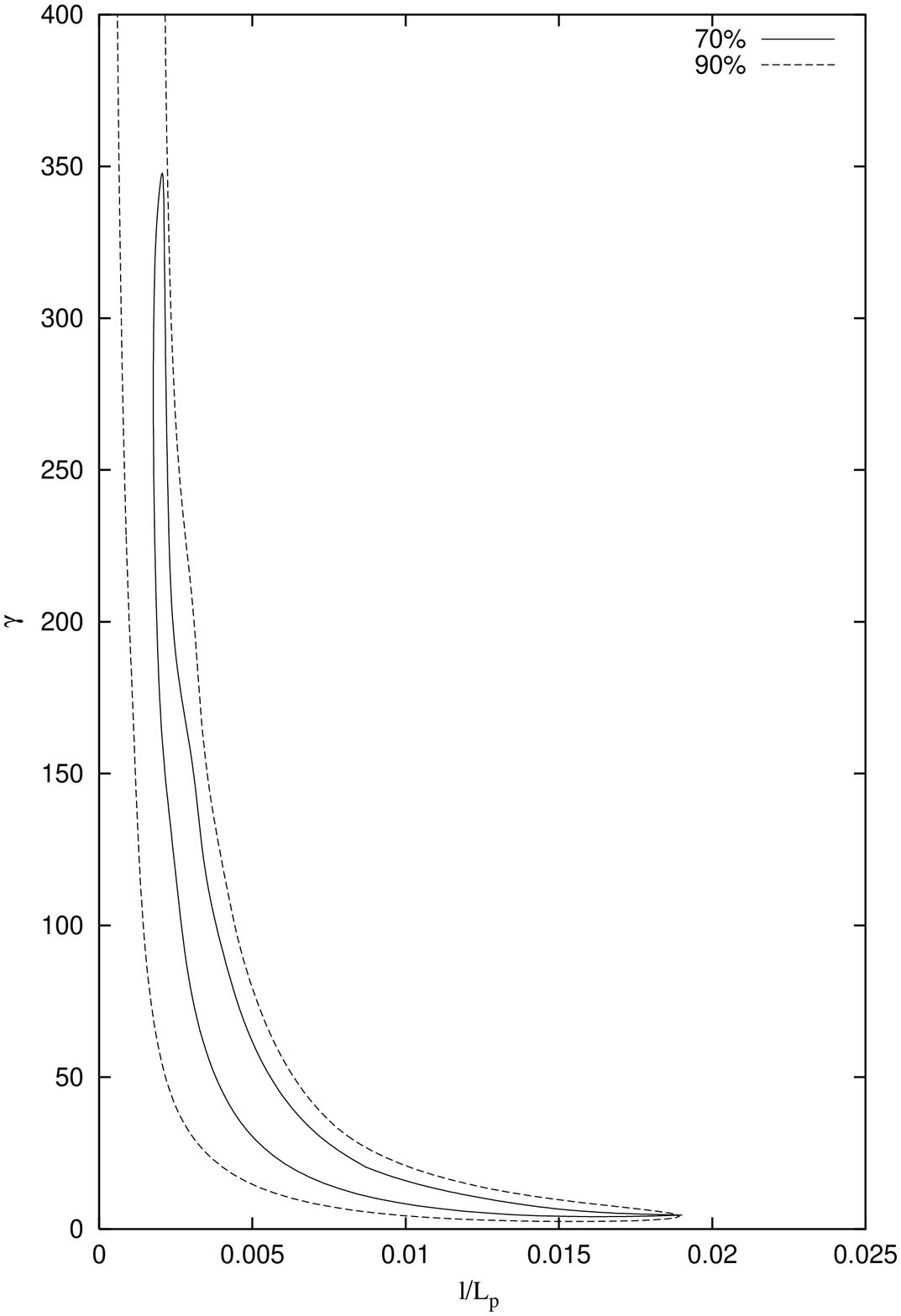}
\hspace{1cm}
\includegraphics[scale=0.4]{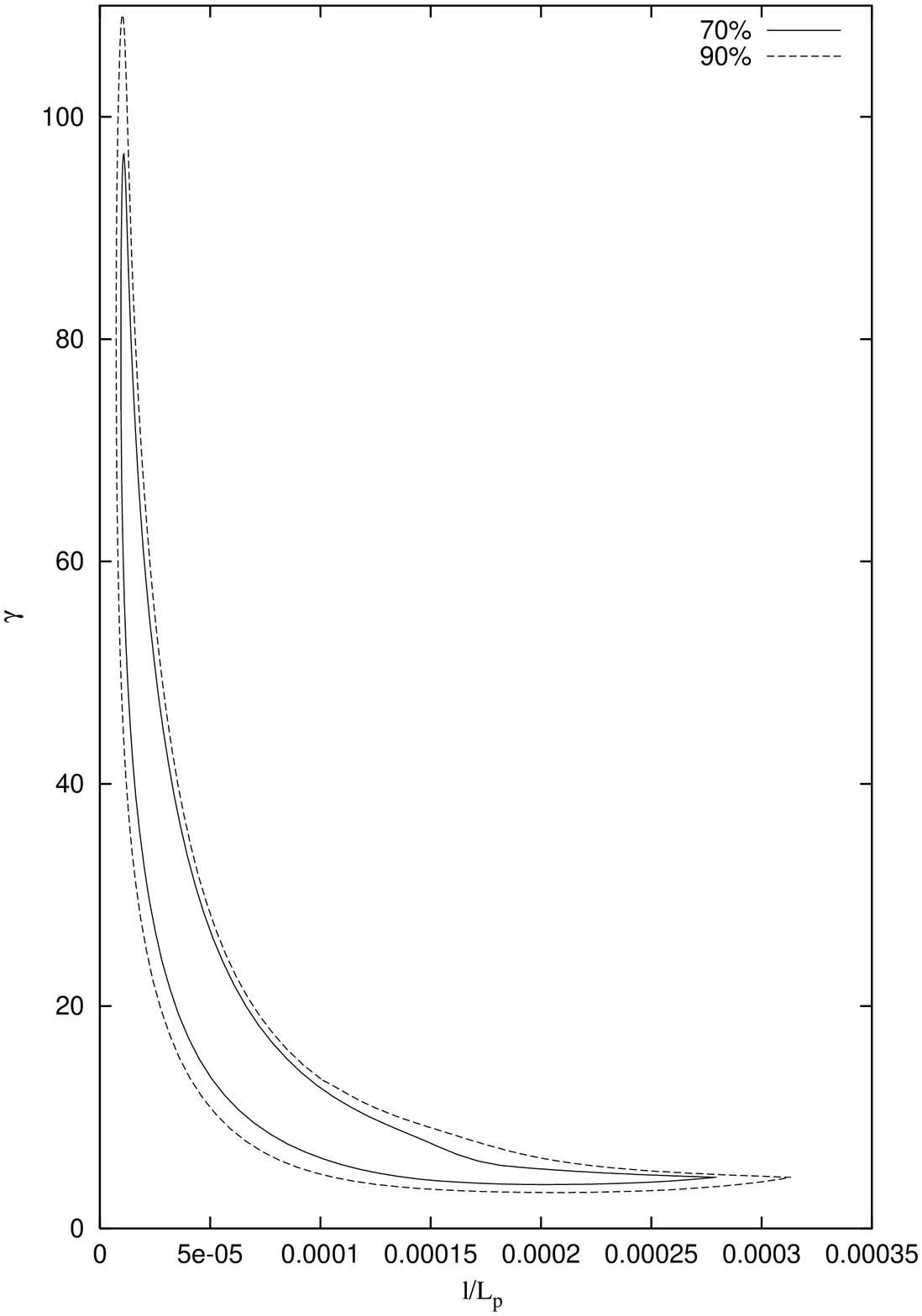}
\end{center}
\caption{ Contours for Beckenstein's models }
\end{figure}

\end{document}